\documentstyle[pre,aps,epsfig,floats]{revtex}
\begin{document}
\draft
\wideabs{
\title{ Multi-Overlap Simulations for Transitions between
Reference Configurations }
\author{Bernd A. Berg$^{1,2}$, Hirochi Noguchi$^{3,*}$ and 
Yuko Okamoto$^{3,4}$}
\address{ (E-mails: berg@csit.fsu.edu, noguchi@ims.ac.jp,
               okamotoy@ims.ac.jp )\\
$^1\,$Department of Physics, Florida State University,
       Tallahassee, FL~32306, USA\\
$^2\,$School of Computational Science and Information Technology\\
Florida State University, Tallahassee, FL~32306, USA\\
$^3\,$Department of Theoretical Studies, Institute for 
      Molecular Science\\ Okazaki, Aichi 444-8585, Japan\\
$^4\,$Department of Functional Molecular Science, Graduate University 
      for Advanced Studies\\ Okazaki, Aichi 444-8585, Japan} 
\date{printed \today}
\maketitle
\begin{abstract}
We introduce a new procedure to construct weight factors, which flatten 
the probability density of the overlap with respect to some pre-defined 
reference configuration. This allows one to overcome free energy barriers
in the overlap variable. Subsequently, we generalize the approach to 
deal with the overlaps with respect to two reference configurations 
so that transitions between them are induced. We illustrate our 
approach by simulations of the brainpeptide Met-enkephalin with the ECEPP/2 
energy function using the global-energy-minimum and the second 
lowest-energy states as reference configurations. The free energy is 
obtained as functions of the dihedral and the root-mean-square distances 
from these two configurations. The latter allows one to identify the 
transition state and to estimate its associated free energy barrier.
\end{abstract}
\pacs{PACS: 05.10.Ln, 87.53.Wz, 87.14.Ee, 87.15.Aa}
}

\narrowtext

\section{Introduction} \label{Introduction}

\footnotetext{$^*\,$Present address: Theory II, Institute of Solid 
State Research, Forschungszentrum J\"ulich, D-52425 J\"ulich, Germany.  
E-mail: hi.noguchi@fz-juelich.de. }
Markov chain Monte Carlo (MC) simulations, for instance by means of the
Metropolis method~\cite{Me53}, are well suited to simulate generalized
ensembles. Generalized ensembles do not occur in nature, but are of 
relevance for computer simulations (see~\cite{HO99,MSO01,Be02} for recent 
reviews). They may be designed to overcome free energy barriers, which 
are encountered in Metropolis simulations of the
Gibbs-Boltzmann canonical ensemble. Generalized ensembles do still 
allow for rigorous estimates of the canonical expectation values,
because the ratios between their weight factors and the canonical 
Gibbs-Boltzmann weights are exactly known. 

Umbrella sampling~\cite{ToVa77} was one of the earliest generalized-ensemble 
algorithms. 
In the multicanonical approach~\cite{BeNe91,BeCe92} one 
weights with a microcanonical temperature, which corresponds, in a 
selected energy range, to a working estimate of 
the inverse density of states.  Expectation values of the canonical 
ensembles can be constructed for a wide temperature range, hence the name 
``multicanonical". Here, ``working estimate'' means that running the 
updating procedure with the (fixed) multicanonical weight factors 
covers the desired energy range. The Markov process exhibits random 
walk behavior and moves in cycles from the maximum (or above) to the
minimum (or below) of the chosen energy range, and back. A working 
estimate of the multicanonical weights allows for calculations of
the spectral density and all related thermodynamical observables 
with any desired accuracy by 
simply increasing the MC statistics. Thus, we have a two-step approach: 
The first step is to obtain the working estimate of the weights, and
the second step is to perform a long production run with these
weights.
There is no need for that estimate to converge towards the exact 
inverse spectral density. Once the working estimate of the weights 
exists, MC simulations with frozen weights converge and allow one to 
calculate thermodynamical observables with, in principle, arbitrary 
precision.  Various methods, ranging from finite-size scaling 
estimates~\cite{BeHaNe93} in case of suitable systems to general 
purpose recursions~\cite{Be96,SuOk00,WaLa01}, are at our disposal to 
obtain a working estimate of the weights.

In the present article we deal with a variant of the 
multicanonical 
approach: Instead of flattening the energy distribution, we construct 
weights to flatten the probability density of the overlap with a given 
reference configuration. This allows one 
to overcome energy barriers in the 
overlap variable and to get accurate estimates of thermodynamic 
observables at overlap values which are rare in the canonical ensemble. 
A similar concept was previously used in spin glass 
simulations~\cite{BeBiJa00}, but there is a crucial difference: In 
Ref.\cite{BeBiJa00} the weighting was done for the self-overlap of 
two replicas of the system and a proper name would be multi-self-overlap 
simulations, while in the present article we are dealing with the overlap 
to a predefined configuration. 

We next generalize our approach to deal with two reference 
configurations so that transitions between them become covered and 
our method allows one then to estimate the transition states and its 
associated free energy barrier. We have in mind situations where 
experimentalists determined the reference configurations
and observed transitions between them, but an understanding of the
free energy landscape between the configurations is missing.
An example would be the conversion from a configuration with 
$\alpha$ helix structures to a native structure which is mostly in
the $\beta$ sheet, as it is the case for 
$\beta$-lactoglobulin~\cite{KYMSN87,HSG96}.

The paper is organized as follows: In the next section we describe
the algorithmic details, using first one and then two reference 
configurations. In particular, a two-step updating procedure
is defined, which is typically more efficient than the conventional
one-step updating. Moreover, based on the sums of uniformly distributed
random numbers, a method to obtain a working estimate of the 
multi-overlap weights is introduced.
In section~\ref{Met-en} we illustrate the method for a simulation with
the pentapeptide Met-enkephalin. Our simulations use the all-atom 
energy function ECEPP/2 (Empirical Conformational Energy Program for 
Peptides~\cite{SNS84}) and rely on its 
implementation in the computer package SMMP (Simple Molecular Mechanics 
for Proteins~\cite{Ei01}). We use as reference configurations the 
global energy minimum (GEM) state, which has been determined by
many authors~\cite{LiSc87,VFBr91,OKK92,HaOk93,MMMV94}, and the second 
lowest-energy state, as identified in Refs.\cite{OKK92,MHO98}.
While our overlap definition relies on a
distance definition in the space of the dihedral angles, it turns
out that for the data analysis the use of the root-mean-square (rms)
distance is crucial. It is only in the latter variable that one obtains 
a clear picture of the transition saddle point in the two-dimensional
free energy diagram.
In the final section a summary of the present results and an 
outlook with respect to future applications are given.

\section{Multi-Overlap Metropolis Algorithm} \label{Multi-Overlap}

In this section we explain the details of our multi-overlap algorithm. 
The 
overlap of a configuration versus a reference configuration is defined 
in the next subsection. In the second subsection we discuss details of 
the updating. To achieve step one of the method, i.e., the construction 
of a working estimate of the multi-overlap weights, one could employ 
a similar recursion as the one used in~\cite{BeBiJa00} or explore 
the approach of~\cite{WaLa01}. Instead of doing so, we decided to 
test a new method: At infinite temperature, $\beta =0$, the overlap 
distributions can be calculated analytically (see 
subsection~\ref{sums_uniform}). We use this as starting point and 
estimate the overlap weights at the desired temperature by increasing 
$\beta$ in sufficiently small steps so that the entire overlap 
range remains covered. In the final subsection we define the overlap 
with respect to two distinct reference configurations to cover
the transition region between them.

\subsection{Definition of the overlap} \label{Overlap}

There is a considerable amount of freedom in defining the overlap 
of two configurations. For instance, one may rely on the rms 
distance between configurations, and in subsection~\ref{phy_res} 
we analyze some of our results 
in this variable. However, the computation of the rms distance
is slow and for MC calculations it is important to rely on a 
computationally fast definition. Therefore, we define the overlap in 
the space of dihedral angles by, as it was already used 
in~\cite{HMO97}, 
\begin{equation} \label{dihedral_ovl}
 q\ =\ (n - d)/n\ ,
\end{equation}
where $n$ is the number of dihedral angles and $d$ is the distance 
between configurations defined by
\begin{equation} \label{dihedral_d}
  d\ =\ ||v-v^1||\ =\ {1\over \pi} \sum_{i=1}^n d_a(v_i,v_i^1)\ .
\end{equation}
Here, $v_i$ is our generic notation for the dihedral angle $i$,
$-\pi < v_i \le \pi$, and $v^1$ is the vector of dihedral angles
of the reference configuration. 
The distance $d_a(v_i,v_i')$ between two angles is defined by
\begin{equation} \label{d_a}
  d_a(v_i,v_i')\ =\ \min (|v_i-v_i'|,2\pi-|v_i-v_i'|)\ .
\end{equation}
The symbol $||.||$ defines a norm in a vector space. 
In particular, the triangle inequality holds
\begin{equation} \label{d_triangle}
||v^1-v^2|| \le ||v^1-v|| + ||v-v^2||\ .
\end{equation}
For a single angle we have
\begin{equation} \label{d_range}
 0\le |v_i-v_i^1| \le \pi\ \Rightarrow\ 0\le d \le n\ .
\end{equation}
At $\beta =0$ (i.e., infinite temperature)
\begin{equation} \label{d_i}
 d_i\ =\ {1\over \pi}\, d_a(v_i,v_i^1)
\end{equation}
is a uniformly distributed random variable in the range 
$0\le d_i\le 1$ and 
the distance $d$ in (2) becomes the sum of $n$ such uniformly 
distributed random variables, which allows for an exact calculation 
of its distribution. 

\subsection{Multi-overlap weights} \label{muov_wghts}

We choose a reference configuration of $n$ dihedral angles 
$v_i^1,\, (i=1,\dots,n)$ to define the dihedral 
distance~(\ref{dihedral_d}). We want to simulate the system with 
weight factors that lead to a random walk (RW) process in 
the dihedral distance $d$, 
\begin{equation} \label{RWC}
d < d_{\min}\ \to\ d> d_{\max} ~~{\rm and~~back}\ . 
\end{equation}
Here, $d_{\min}$ is chosen sufficiently small so that one can
claim that the reference configuration has been reached, e.g., a
few percent of $n/2$, which is the average $d$ at
$T=\infty$. The value of $d_{\max}$ has to be sufficiently large 
to introduce a considerable amount of disorder, e.g., 
$d_{\max}=n/2$.  In the following we call one event of the 
form~(\ref{RWC}) a random walk cycle (RWC).

One possibility is to choose weight factors which give a flat 
probability density in the dihedral distance range $0\le d\le n/2$, 
falling off for $d>n/2$ by keeping the $d$-dependence of the weight 
constant for $d\ge n/2$. This is quite similar to multimagnetical
simulations~\cite{BeHaNe93}, for which the external magnetic field 
takes the place of the reference configuration. The analogy becomes 
obvious, when the external field is defined via a ghost spin, which 
couples to all other spins. For instance, the spins $\vec{s}$ of the 
Heisenberg ferromagnet are three-dimensional vectors of magnitude 
$\vec{s}^{\,2}=1$.  Their interaction with an external magnetic 
field $\vec{H}$ can be written as
\begin{equation} \label{Hspin}
 \vec{H}\cdot\sum_i \vec{s}_i = H\sum_i \vec{s}_H\cdot\vec{s}_i = 
N\,H\,q\ ,
\end{equation}
where $\vec{s}_H$ is the unit vector in the direction of the magnetic
field, $\vec{s}_i$ is the Heisenberg spin at site $i$, $N$ is the number
of spins, and $q$ is the overlap of the spin configuration with the
reference configuration $\vec{s}_H$:
\begin{equation} \label{q_spin}
 q = {1\over N} \sum_i \vec{s}_H\cdot\vec{s}_i\ . 
\end{equation}
Using the multi-overlap language~\cite{BeBiJa00}, the 
multi-mag\-neti\-cal~\cite{BeHaNe93} weight factors may 
then be re-written as
\begin{equation} \label{q_wghts}
\exp \left( - \beta E + S(q) \right) = w_c(E)\,w_q(q)~,
\end{equation}
where
\begin{equation} \label{w_c}
w_c(E)=\exp (-\beta\,E)~,
\end{equation}
and $E=-\sum_{\langle ij\rangle}\vec{s}_i\cdot\vec{s}_j$
is energy function of the Heisenberg ferromagnet (the sum is over
nearest neighbor spins).  Here, $S(q)$ has the meaning of a microcanonical 
entropy of the overlap parameter, which has to be determined so that 
the probability density becomes flat in $q$. Weights for other than the 
flat distribution have also been discussed in the literature, e.g.,
Ref.\cite{HeSt95}, on which we shall comment in connection with
figure~\ref{fig_RWCs} below.

\subsection{The updating procedure} \label{updating}

In essence, there are two ways to implement the update.

\begin{enumerate}

\item Combine the multi-overlap and the canonical weights to one
probability, which is accepted or rejected in one random step.

\item Accept or reject the multi-overlap and the canonical
probabilities sequentially in two random steps.

\end{enumerate}

\subsubsection{One-step updating}

As defined in equations~(\ref{q_wghts}) and~(\ref{w_c}), the 
weight factor is a product of 
$w_c(E)$ and $w_q(d)$, where $w_c(E)$ is the usual, canonical 
Gibbs-Boltzmann factor and $w_q(d)$ is the multi-overlap weight 
factor, where we now use the distance $d$ from the reference 
configuration (instead of the overlap $q$) as argument. As is 
clear from 
equation~(\ref{dihedral_ovl}), the use of either $q$ or $d$ as 
argument is equivalent, while in the presentation of results the 
use of either variable can have intuitive advantages. In the 
one-step updating we combine the weights to
\begin{equation} \label{wproduct} 
w(E,d)\ =\ w_c(E)\, w_q(d)~,
\end{equation}
and accept or reject newly proposed configurations in the standard 
Metropolis way. Notably, the calculation of $w_q(d)$ (a simple table 
lookup) is very fast compared with the calculation of $w_c(E)$. 
Therefore, the following two-step procedure is of interest.

\subsubsection{Two-step updating}

Suppose that the present configuration is $(d,E)$ and a new 
configuration $(d',E')$ is proposed:
\begin{equation} \label{to_prime} 
(d,E)\ \to\ (d',E')\ . 
\end{equation}
We can sequentially first accept or reject with the $w_q(d)$ 
probabilities and then conditionally, when the $d$-part is 
accepted, with the $w_c(E)$ probabilities.

Proof: We show detailed balance for two subsequent updates of the same 
dihedral angle with the two-step procedure. There are four cases
with probabilities of acceptance:
\begin{equation} \label{to_prime_probs} 
P_i ,\ i=1,2,3,4\,.
\end{equation}
They are listed in the following:
\begin{eqnarray} \nonumber
1.~~& & w_q(d') \ge w_q(d) ~~{\rm and}~~ w_c(E') \ge w_c(E):\\
    & & P_1 = 1 , \\ \nonumber
2.~~& & w_q(d') \ge w_q(d) ~~{\rm and}~~ w_c(E') <   w_c(E):\\
    & & P_2 = w_c(E')/w_c(E) , \\ \nonumber
3.~~& & w_q(d') <   w_q(d) ~~{\rm and}~~ w_c(E') \ge w_c(E):\\
    & & P_3 = w_q(d')/w_q(d) , \\ \nonumber
4.~~& & w_q(d') <   w_q(d) ~~{\rm and}~~ w_c(E') <   w_c(E):\\
    & & P_4 =  w_q(d')\,w_c(E') / [ w_q(d)\,w_c(E) ] .
\end{eqnarray}
For the inverse move
\begin{equation} \label{from_prime} 
(d',E')\ \to\ (d,E) 
\end{equation}
with probabilities of acceptance
\begin{equation} \label{from_prime_probs} 
P'_i ,\ i=1,2,3,4,
\end{equation}
the cases are:
\begin{eqnarray}  \nonumber
1.~~& & w_q(d) \le w_q(d') ~~{\rm and}~~ w_c(E) \le w_c(E'):\\
    & & P'_1 = w_q(d)\,w_c(E) / [ w_q(d')\,w_c(E') ] ,\\ \nonumber
2.~~& & w_q(d) \le w_q(d') ~~{\rm and}~~ w_c(E) > w_c(E'):\\
    & & P'_2 = w_q(d)/w_q(d') , \\ \nonumber
3.~~& & w_q(d) > w_q(d') ~~{\rm and}~~ w_c(E) \le w_c(E'):\\
    & & P'_3 = w_c(E)/w_c(E') , \\ \nonumber
4.~~& & w_q(d) > w_q(d') ~~{\rm and}~~ w_c(E) > w_c(E'):\\
    & & P'_4 = 1 .
\end{eqnarray}
For the ratios we find 
\begin{equation} \label{ratios} 
{P_i\over P'_i}\ =\ {w_q(d')\,w_c(E')\over w_q(d)\,w_c(E)}~,
\end{equation}
independently of $i=1,2,3,4$. Therefore, we have constructed
a valid Metropolis updating procedure.

\subsection{Sums of a uniformly distributed random variable}
\label{sums_uniform}

To calculate the overlap weights at infinite temperature, 
we consider the sum 
\begin{equation} \label{uniform_n}
 u^r\ =\ x^r_1 + \dots + x^r_n
\end{equation}
of the random variables $x^r_j$ ($j=1, \cdots, n$), 
each uniformly distributed in the 
interval $[0,1)$ and derive a recursion formula for the probability 
density $f_n(u)$ of this distribution. Care is taken to cast the 
recursion in a form which allows for a numerically stable 
implementation~\cite{BeBook} over a reasonably large range of $n$.

Let us recall the probability density of the uniform 
distribution:
\begin{equation} \label{uniform_1}
 f_1 (x)\ =\ \cases{ 1, ~{\rm for}~ 0 \le x < 1, \cr
                       0, ~{\rm otherwise}. }
\end{equation}
To derive the recursion formula for the probability density of 
the random variable~(\ref{uniform_n}), it is convenient to cast 
it in the form
\begin{equation} \label{uniform_fn}
 f_n (u)\ =\ \sum_{k=1}^n f_{n,k} (x_k) ~~{\rm with}~~\
 x_k = u-k+1 , 
\end{equation}
where
\begin{equation}
 f_{n,k} (x)\ =\ \cases{\displaystyle{\sum_{i=0}^{n-1} a_{n,k}^i\, x^i},~{\rm for}~
                                0\le x < 1\\ \label{uniform_fnk},\cr
                           ~0,~{\rm otherwise}. } 
\end{equation}
The master formula for the recursion is obtained from the 
convolution
\begin{equation} \label{u_convo}
 f_n (u)\ =\ \int_0^u f_1 (u-v)\ f_{n-1} (v)\ dv\ .
\end{equation}
Let now $u=x+k-1$ with $0\le x< 1$, and equations~(\ref{uniform_1}),
(\ref{uniform_fn}), and (\ref{uniform_fnk}) imply
\begin{eqnarray} \nonumber
 f_{n,k} (x)\ =\ \int_{k-2+x}^{k-1+x} f_{n-1} (v)\ dv\\
 =\ \int_x^1 f_{n-1,k-1} (y)\ dy\
+\ \int_0^x f_{n-1,k} (y)\ dy\ . \label{ufnk_integral}
\end{eqnarray}
Using equation~(\ref{uniform_fnk}) and performing the 
integrations, we obtain
\begin{eqnarray} \nonumber 
f_{n,k} (x) &=& \sum_{i=0}^{n-2} a_{n-1,k-1}^i\,{1 \over i+1}\ 
   -\  \sum_{i=0}^{n-2} a_{n-1,k-1}^i\,{x^{i+1}\over i+1}\\
  &+&  \sum_{i=0}^{n-2} a_{n-1,k}^i\,  {x^{i+1}\over i+1}\ .
\label{ufnk_recursion}
\end{eqnarray}
Expanding in powers of $x$ and comparing~(\ref{uniform_fnk}) 
with~(\ref{ufnk_recursion}) allows one to calculate the coefficients 
$a_{n,k}^i$ recursively in a numerically robust way:
\begin{equation} \label{anki_recursion} 
a_{n,k}^0 = \sum_{j=0}^{n-1} {a_{n-1,k-1}^j\over j+1}\,,\
a_{n,k}^i = \sum_{j=0}^{n-1} {a_{n-1,k}^j-a_{n-1,k-1}^j\over j+1}\ .
\end{equation}
Once the coefficients $a_{n,k}^i$ are available, one can easily 
evaluate the probability densities $f_n (u)$ and the corresponding 
cumulative distribution functions.

The probability density~(\ref{uniform_fn}) takes its maximum value
for $u=n/2$. Due to the central limit theorem the fall-off behavior
is Gaussian as long as $u$ stays sufficiently close to $n/2$. In 
the tails, for $u\to 0$ or $u\to n$, the fall-off is much faster
than Gaussian, namely an exponential of an exponential as follows
from extreme value statistics~\cite{Gumb}.

\subsection{Combination of two weights}

In the following the weights with superscript $j$, $w_q^j(d_j)$, 
correspond to two distinct reference configurations $v^j$, ($j=1,2$), 
and $d_j$ is the distance from the configuration at hand to the 
configuration $v^j$.
Let us assume that multi-overlap simulations with respect to
the two reference configurations have been carried out and that
the weights, $w_q^1(d_1)$ and $w_q^2(d_2)$, have been determined 
so that they sample their distance distributions approximately 
uniformly. 

We want to construct combined weights $w_q^{12}(d_1,d_2)$ which 
lead to a RW process between the configurations 
$v^1$ and $v^2$. Our choice is
\begin{equation} \label{w12}
w_q^{12}(d_1,d_2)\ =\ \cases{ 
                w_q^1(d_1),~~{\rm for}~~d_1<d_2\,, \cr
                c_j\,w_q^2(d_2),~~{\rm for}~~ d_1\ge d_2\,.} 
\end{equation}
The constant $c_j$, with $j$ either 1 or 2, is introduced to allow 
for smooth transitions from $d_1 < d_2$ to $d'_1 \ge d'_2$ and vice 
versa. We determine $c_j$ from the analysis of either run~1 (or run~2), 
which are the (one reference configuration) simulations leading to the 
weights $w_q^1(d_1)$ (or $w_q^2(d_2)$).
The constant $c_1$ is found from run 1 by scanning the time series for 
configuration for which $d_1\ge d_2$ holds and which have a one-update 
transition $(d_1,d_2)\to (d'_1,d'_2)$ with $d'_1<d'_2$. From these 
configurations $k$ we determine the constant $c_1$ so that
\begin{equation} \label{c1}
\sum_k w_q^1[d_1(k)]\ =\ c_1 \sum_k w_q^2[d_2(k)] 
\end{equation}
holds. Similarly, run 2 may be used to get $c_2$. It turns out that
the normalized weights almost agree in the transition region
and, therefore, the patching (\ref{w12}) works. The dependence of
the constant on the run used for its determination is small, and 
it appears not worthwhile to explore more sophisticated methods.

\begin{figure}[t] \begin{center}
\epsfig{figure=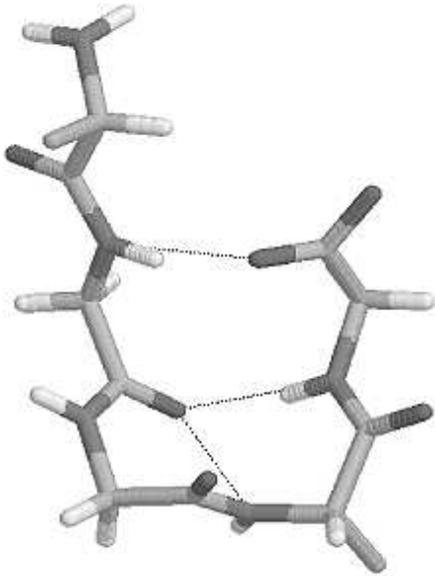,width=6.0cm} \vspace*{5mm}
\caption{Reference configuration~1.  Only backbone structure is shown.
The N-terminus is on the left-hand side and the C-terminus on the
right-hand side.  The dotted lines stand for hydrogen bonds.
The figure was created with RasMol [23] \label{fig_A4} }
\end{center} \end{figure}

It is straightforward to implement the Metropolis updating with 
respect to the weights~(\ref{w12}). For the transition
\begin{equation} \label{d1d2}
(d_1,d_2)\ \to\ (d'_1,d'_2)\,,
\end{equation}
one has to distinguish four more cases:
\begin{eqnarray}
1. & & d_1 <   d_2 ~~{\rm and}~~ d'_1 <   d'_2\,,\\
2. & & d_1 <   d_2 ~~{\rm and}~~ d'_1 \ge d'_2\,,\\
3. & & d_1 \ge d_2 ~~{\rm and}~~ d'_1 <   d'_2\,,\\
4. & & d_1 \ge d_2 ~~{\rm and}~~ d'_1 \ge d'_2\,.
\end{eqnarray}

Alternatively to the approach outlined, one may combine $d_1$ and
$d_2$ into a new variable $\theta_d$ for which the weights are then 
calculated as in the one-dimensional case. A suitable choice along 
this line is
\begin{equation} \label{theta_d1d2}
\theta_d = {2\over\pi}\, \arctan \left( {d_1\over d_2}\right)\ .
\end{equation}

\section{Met-Enkephalin Simulations} \label{Met-en}

In the following we introduce two reference configurations.
Subsequently, we discuss first the results for simulations  
with one reference configuration and then those involving 
both reference configurations.

\subsection{The reference configurations}
 
\begin{figure}[t] \begin{center}
\epsfig{figure=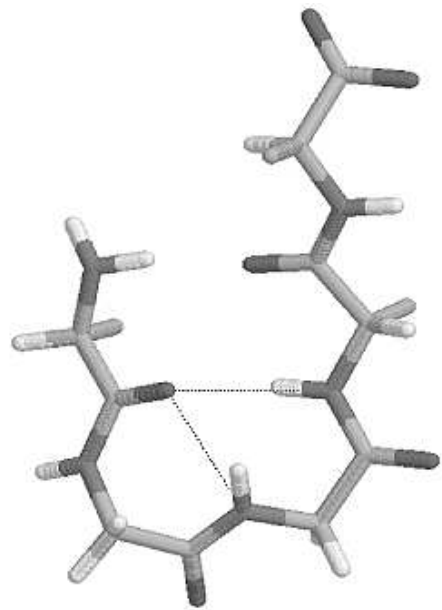,width=6.0cm} \vspace*{5mm}
\caption{Reference configuration~2.  See the caption of
figure 1 for details. \label{fig_B4} }
\end{center} \end{figure}

Met-enkephalin has the amino-acid sequence Tyr-Gly-Gly-Phe-Met.
We fix the peptide-bond dihedral angles $\omega$ to $180^{\circ}$, which implies
that the total number of variable dihedral angles is $n=19$. We
neglect the solvent effects as in previous works. The low-energy
configurations of Met-enkephalin in the gas phase have been 
classified into several groups of similar structures~\cite{OKK92,MHO98}. Two
reference configurations, called configuration~1 and 
configuration~2, are used in the following and depicted in
figures~\ref{fig_A4} and~\ref{fig_B4}, respectively. 
Configuration~1 has a $\beta$-turn structure
with hydrogen bonds between Gly-2 and Met-5, and configuration~2 a
$\beta$-turn with a hydrogen bond between Tyr-1 and Phe-4~\cite{MHO98}. 

For our present work the two reference configurations were obtained 
by minimizing the GEM and the second lowest energy state of previous
literature with respect to the ECEPP/2 energy function. The 
minimization was performed with the SMMP minimizer~\cite{Ei01} and 
by quenching.  Both methods gave identical final energies. 
In table~\ref{tab_ref_conf}
we list the variable dihedral angles of the configurations before and 
after this minimization. The initial dihedral angles for the GEM are
taken from table~1 of Ref.\cite{MMMV94} and the initial dihedral
angles for the second lowest energy state B are from table~1
of Ref.\cite{OKK92}. In table~\ref{tab_ref_conf} we give the angles 
in degrees, while for the MC simulations radians were used as in
equations~(\ref{dihedral_ovl}) and~(\ref{dihedral_d}) for the overlap. 
Our labeling of the residues follows the SMMP convention and deviates
from those of Refs.\cite{MMMV94,OKK92}.

\begin{table}[t]
\caption{Met-enkephalin reference configurations. The columns 
$\rm GEM_{min}$ and $\rm B_{min}$ correspond to configuration~1 
and configuration~2, respectively.  \label{tab_ref_conf}}
\vspace{2mm}
\centering
\begin{tabular}{|c|c|c|c|c|c|}                       
Residue & Angle & GEM~\cite{MMMV94} 
                           &$\rm GEM_{min}$
                                      & B~\cite{OKK92}        
                                               &$\rm B_{min}$
                                                          \\ \hline
 1    &$\chi_1$ & $-179.9$ & $-179.8$ & $-179$ & $+179.4$ \\ \hline
 1    &$\chi_2$ & $-111.3$ & $-111.4$ & $~-95$ & $~-94.3$ \\ \hline
 1    &$\chi_6$ & $+145.3$ & $+145.3$ & $+169$ & $-179.9$ \\ \hline
 1      &$\phi$ & $~-86.4$ & $~-86.3$ & $+111$ & $~+55.7$ \\ \hline
 2      &$\psi$ & $+153.7$ & $+153.7$ & $+157$ & $+157.6$ \\ \hline
 2      &$\phi$ & $-161.6$ & $-161.5$ & $~-71$ & $~-70.7$ \\ \hline
 3      &$\psi$ & $~+71.2$ & $~+71.1$ & $~+78$ & $~+78.0$ \\ \hline
 3      &$\phi$ & $~+64.1$ & $~+64.1$ & $~159$ & $+156.5$ \\ \hline
 4      &$\psi$ & $~-93.5$ & $~-93.5$ & $~-37$ & $~-35.7$ \\ \hline
 4    &$\chi_1$ & $+179.8$ & $+179.8$ & $~+59$ & $~+55.3$ \\ \hline
 4    &$\chi_2$ & $~+80.0$ & $~+80.0$ & $~+87$ & $~+86.8$ \\ \hline
 4      &$\phi$ & $~-81.7$ & $~-81.7$ & $-154$ & $-155.7$ \\ \hline
 5      &$\psi$ & $~-29.2$ & $~-29.2$ & $+151$ & $+151.6$ \\ \hline
 5    &$\chi_1$ & $~-65.1$ & $~-65.1$ & $~-68$ & $~-69.4$ \\ \hline
 5    &$\chi_2$ & $-179.2$ & $-179.2$ & $+177$ & $-176.3$ \\ \hline
 5    &$\chi_3$ & $-179.3$ & $-179.3$ & $-179$ & $-179.7$ \\ \hline
 5    &$\chi_4$ & $~-60.0$ & $~-59.9$ & $~+60$ & $~+59.9$ \\ \hline
 5      &$\phi$ & $~-80.8$ & $~-80.7$ & $-140$ & $-140.0$ \\ \hline
 5    &$\psi_t$ & $+143.9$ & $+143.5$ & $~-29$ & $~-30.6$ \\
\end{tabular}
\end{table}

The distance between the two minimized configurations is $d=6.62$ 
($q=0.652$) and their energies are given in table~\ref{tab_Met_emin}.

\begin{table}[ht]
\centering
\caption{ Energies (in kcal/mol) of the Met-enkephalin reference
configurations 1 and~2.  \label{tab_Met_emin}}
\vspace{2mm}
\begin{tabular}{|c|c|c|c|c|c|}                       
  & Total    & Coulomb & Lennard-Jones & H-Bond   & Torsion \\ \hline
1 & $-$10.72 & $+$21.41   & $-$27.10      & $-$6.21  & $+$1.19 \\ \hline
2 & $~-$8.42 & $+$22.59   & $-$26.38      & $-$4.85  & $+$0.23 \\
\end{tabular}
\end{table}

\subsection{Simulations with one reference configuration}

Each of our multi-overlap simulations at fixed temperature relies on
a statistics of 16,777,216 sweeps for which data are recorded in a
time series of 524,288 events, i.e., with a stepsize of 32 sweeps. 
We started most of our simulations with the GEM configuration, but
some random starts were also performed and no noticeable differences
were encountered.

\begin{figure}[t] \begin{center}
\epsfig{figure=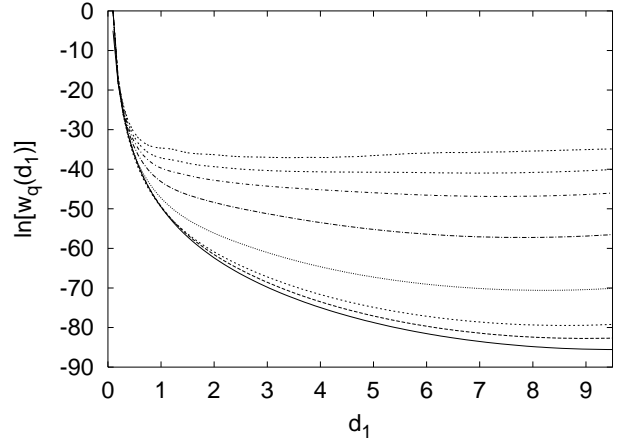,width=\columnwidth} \vspace*{0mm}
\caption{Weight estimates from simulations with reference
configuration~1. From up to down the weight functions correspond 
to the following temperatures: $230\,$K, $300\,$K, $400\,$K, 
$700\,$K, $2,000\,$K, $10,000\,$K, $100,000\,$K and infinity 
($\beta =0$).  \label{fig_weights1} }
\end{center} \end{figure}

Starting with the analytical result~(\ref{uniform_fn}), valid at 
$\beta =0$, the weights are calculated by increasing $\beta$ (i.e., 
decreasing the temperature) between simulations slowly so that 
the RW of each simulation still covers the desired overlap range 
when using the weight estimates from the previous temperature. 
Discretization errors due to histograming can be severe and instead of 
weights which are piecewise constant within each one histogram 
interval, we used the interpolation of Ref.~\cite{BeNe91}:
\begin{equation} \label{interpolation}
\ln w(d) = (1-\alpha) \ln w(d_i) + \alpha \ln w(d_{i+1})~, 
~{\rm for}~ d_i \le d < d_{i+1}~,
\end{equation}
where
\begin{equation}
\alpha= \frac{d-d_i}{d_{i+1}-d_i}~.
\end{equation}
Figure~\ref{fig_weights1} depicts the thus obtained weight function 
estimates from simulations with reference configuration~1. 
After five simulations we arrive at the physical temperature 
$T=300\,$K. The same iteration works with reference configuration~2.

For the values $d_{\min}=0.025\,n$ and $d_{\max}=0.495\,n$, where 
$n=19$ is the number of angels in~(\ref{dihedral_d}), we list in
table~\ref{tab_RWC} the number of RWCs~(\ref{RWC}) achieved at each 
temperature. We also list the CPU time ratios for the 1-step 
versus the 2-step updating procedures, which we discussed in the 
previous section. Especially at high temperatures, which are needed
in our approach, the 2-step updating turns out to be more efficient
than the 1-step updating and all of our production runs were done 
with it.

\begin{table}[ht]
\caption{Number of random walk cycles in the simulations with our two reference 
configurations. The last column lists the CPU time ratios for 1-step 
versus 2-step updating. \label{tab_RWC} }
\vspace{2mm}
\centering
\begin{tabular}{|c|c|c|c|}
$T$         & Configuration~1 & Configuration~2 & 1-step/2-step\\ \hline
$100,000\,$K&    9,458        &   9,514         & 3.0 \\ \hline
$ 10,000\,$K&    3,122        &   3,149         & 1.8 \\ \hline
$  2,000\,$K&    2,893        &   2,741         & 1.6 \\ \hline
$    700\,$K&    2,169        &   2,227         & 1.5 \\ \hline
$    400\,$K&    1,342        &   1,693         & 1.3 \\ \hline
$    300\,$K&      462        &     610         & 1.2 \\ \hline
$    230\,$K&       46        &      41         & 1.2 \\
\end{tabular}
\end{table}

We next rely on the peaked distribution function~\cite{BeBook}
to visualize some of the data kept in the time series of our
simulations. The peaked distribution function of a probability
density $f(x)$ is defined by
\begin{equation} \label{Fpeaked}
F_{\rm peaked} (x) = \cases{ F(x) ~~{\rm for}~~ x\le 0.5\, ,\cr
                         1 - F(x) ~~{\rm for}~~ x > 0.5\, ,}
\end{equation}
where 
\begin{equation} \label{F}
F(x) = \int_{-\infty}^x dx'\, f(x')
\end{equation}
is the usual cumulative distribution 
function (see for instance~\cite{PrFl86}).

\begin{figure}[t] \begin{center}
\epsfig{figure=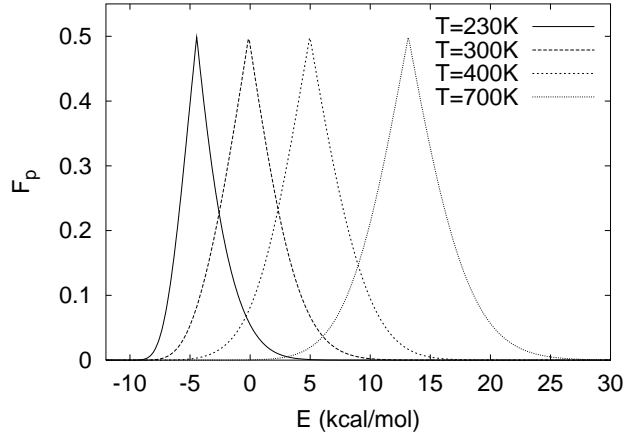,width=\columnwidth} \vspace*{0mm}
\caption{Canonical, peaked energy distributions obtained by
re-weighting multi-overlap simulations. From left to right 
the temperatures used are: $230\,$K, $300\,$K, $400\,$K, 
and $700\,$K. \label{fig_energies1} }
\end{center} \end{figure}

\begin{figure}[t] \begin{center}
\epsfig{figure=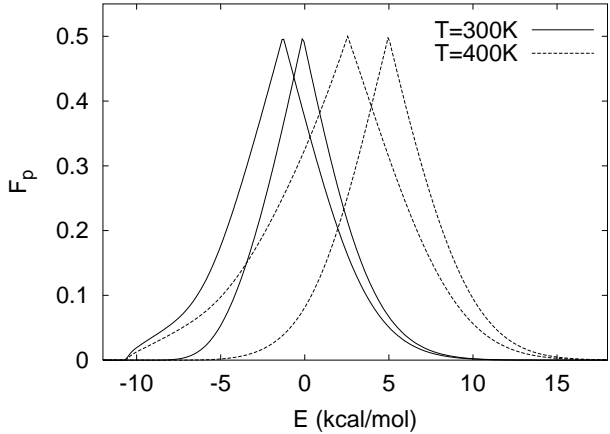,width=\columnwidth} \vspace*{0mm}
\caption{Peaked multi-overlap (left-shifted) and canonical 
energy distributions at $T=300\,$K and 
$T=400\,$K. \label{fig_en1_ovl} }
\end{center} \end{figure}

To visualize how the canonical energy distribution moves when we 
lower the temperature, we plot in figure~\ref{fig_energies1} the 
peaked energy distributions as obtained by re-weighting some of 
the multi-overlap simulations of figure~\ref{fig_weights1} to the 
canonical ensemble of their simulation temperature. Due to the 
re-weighting the distributions look precisely as one expects for 
energies from canonical MC simulations. 
In contrast to conventional canonical simulations, the raw data 
feature a considerably larger number of events at low energies. This 
is illustrated in figure~\ref{fig_en1_ovl}, where we plot the 
$300\,$K and $400\,$K peaked distribution functions of 
figure~\ref{fig_energies1} together with their raw multi-overlap 
peaked distributions 

\begin{figure}[t] \begin{center}
\epsfig{figure=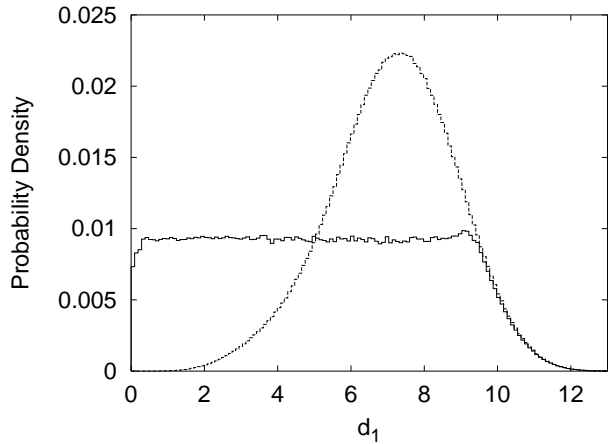,width=\columnwidth} \vspace*{0mm}
\caption{Probability density of the distance from a
multi-overlap simulation at $T=400\,$K (flat) and its 
canonically re-weighted probability density (peaked).  
\label{fig_hist1} }
\end{center} \end{figure}

\begin{figure}[t] \begin{center}
\epsfig{figure=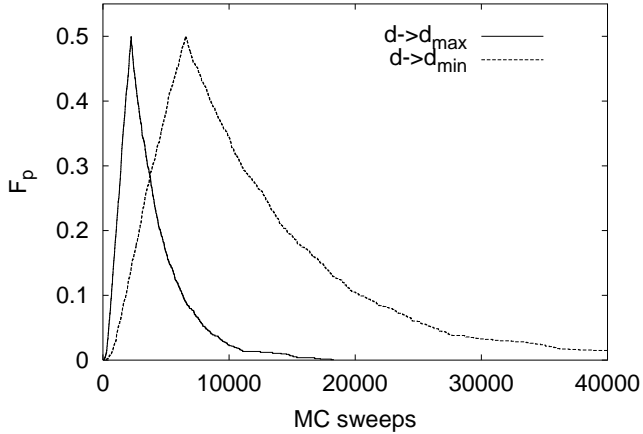,width=\columnwidth} \vspace*{0mm}
\caption{Peaked distribution functions for the forward 
($d\to d_{\max}$) and backward ($d\to d_{\min}$) parts of the
random walk cycles from a multi-overlap simulation at $T=400\,$K. 
\label{fig_RWCs} }
\end{center} \end{figure}

\begin{figure}[t] \begin{center}
\epsfig{figure=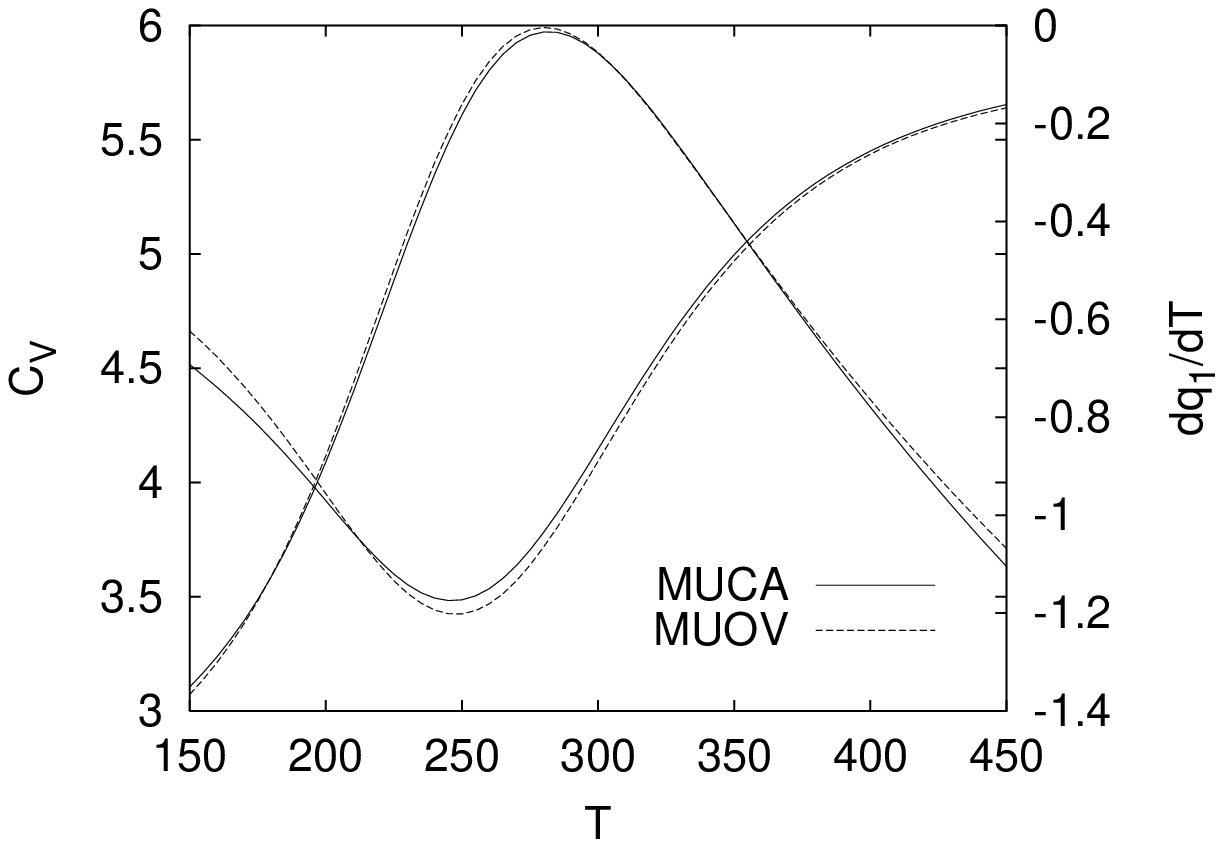,width=\columnwidth} \vspace*{0mm}
\caption{Left-hand-side ordinate: Specific heat re-weighted 
from a multicanonical (MUCA) and from a $300\,$K mul\-ti\--overlap 
(MUOV) simulation with reference configuration~1. Right-hand-side 
ordinate: ${dq_1\over dT}$ re-weighted from the same simulations,
where $q_1$ is the overlap with reference configuration~1. 
\label{fig_cv} }
\end{center} \end{figure}

In figure~\ref{fig_hist1} we give an example of the probability 
density of the distance. For the $400\,$K simulation with reference 
configuration~1 we plot the probability density of $d_1$ as 
obtained from the multi-overlap simulation together with its 
canonically re-weighted probability density. The simulation
itself is run with the multi-overlap weights from the $700\,$K
simulations and the multi-overlap histogram shown is re-weighted
to the multi-overlap $400\,$K weights. As expected, we have 
a flat distribution between 0 and $n/2=9.5$. Moreover, there is
a good coverage of configurations close to the GEM, which are
highly suppressed in the $400\,$K canonical ensemble. The maximum 
ratio of the multi-overlap density divided by the canonical 
density is $6\times 10^{16}$ in this plot.

For the same simulation figure~\ref{fig_RWCs} depicts separately
the peaked distribution function of the forward and backward 
RWCs~(\ref{RWC}). A considerable asymmetry is noticeable and it
turns out that the weights of the 1/k ensemble~\cite{HeSt95}
lead to more RWCs than the flat distribution of 
figure~\ref{fig_hist1}. In connection with our simulations this
is a lucky circumstance, because the 1/k distribution of weights
is in essence the distribution at a somewhat higher temperature
than that of the simulation. This increases the flexibility when 
estimating good weights at a lower temperature from the already 
existing simulation results at a higher temperature. 

For multi-overlap simulations the re-weighting towards low 
temperatures can work much better than for canonical simulations. 
This is due to the fact that the low-energy configurations close 
to low-energy reference configuration are already in the ensemble. 
This is illustrated in figure~\ref{fig_cv}, where we re-weight
the data from a multi-overlap simulation with reference 
configuration~1 at $T=300\,$K and compare with a conventional
multicanonical simulation based on the SMMP package~\cite{Ei01}.
The specific heat $C_V$ and the derivative of the overlap with
respect to the temperature are shown. From $200\,$K to
$400\,$K the deviations of the results are of the order of
the statistical errors, which are not shown for clarity of the
figure. Below $200\,$K deviations of the re-weighted overlap
simulation from the correct behavior become visible, first
in ${dq_1\over dT}$ then in $C_V$. Such deviations are expected 
as the low-energy attractor does not lead to a uniform coverage 
of all low-energy states. The successful re-weighting from 
high simulation temperatures to lower temperatures is an 
improvement, because the Metropolis dynamics at high temperatures 
is faster. But the re-weighting of a multi-overlap simulation
to a lower temperature will fail at some point, because the reference 
configuration introduces a bias towards particular low-energy 
configurations.

The temperature at which $C_V$ and $-{dq_1\over dT}$ take peak
values correspond to the coil-globule transition temperature 
$T_{\theta}$ and the folding temperature $T_f$~\cite{HMO97}.
From figure~\ref{fig_cv} we read off the following
approximate values:
\begin{equation} \label{Ts}
T_{\theta}=280\,{\rm K}~~{\rm and}~~ T_f=245\,{\rm K}\ .
\end{equation}

\subsection{Simulations with two reference configurations}

At $300\,$K we combine the weights from the runs with reference 
configurations~1 and~2 to one weight function according to our
equation~(\ref{w12}). We record now three different RWCs:

\begin{enumerate}

\item With respect to reference configuration 1 from $d_{\min}$ 
to $d_{\max}$ and back, found 315 times.

\item With respect to reference configuration 2 from $d_{\min}$ 
to $d_{\max}$ and back, found 545 times.

\item From $d_{\min}$ of reference configuration 1 to $d_{\min}$ 
of reference configuration 2 and back, found 196 times.

\end{enumerate}

\begin{figure}[t] \begin{center}
\epsfig{figure=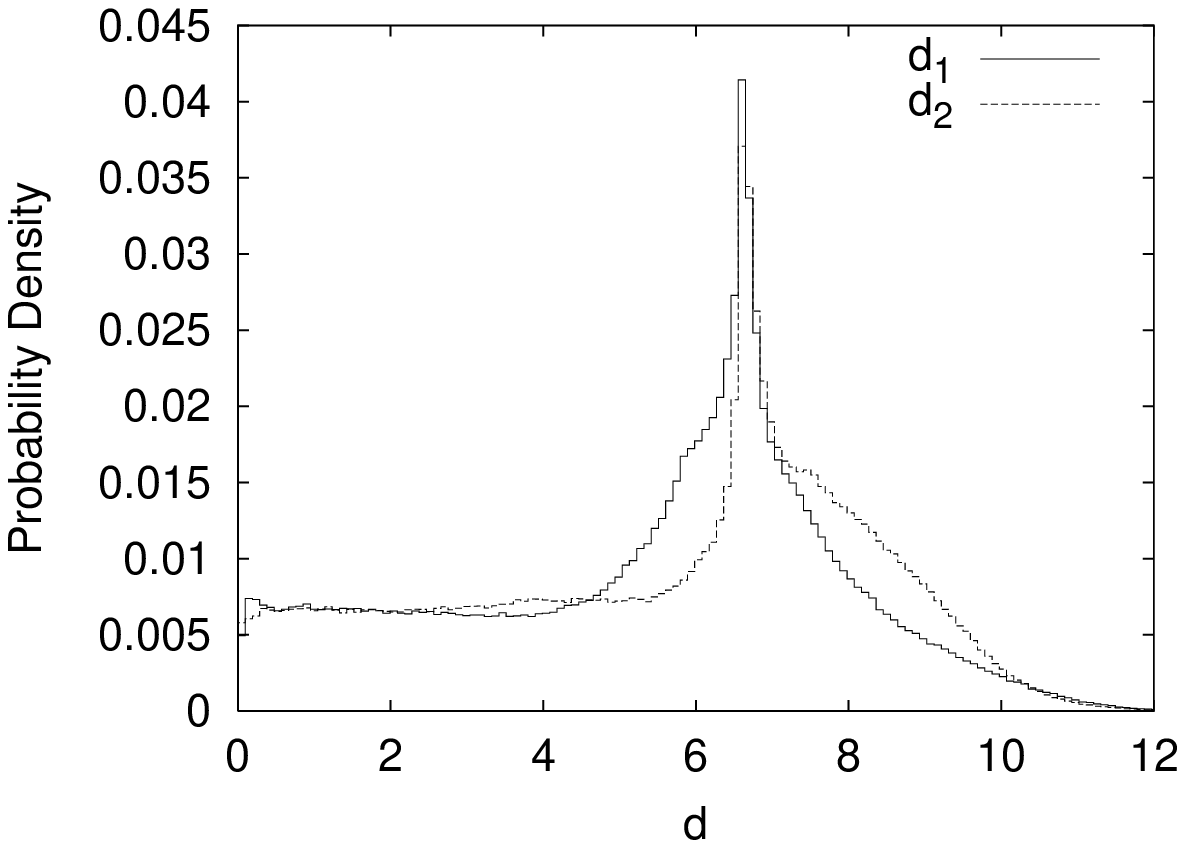,width=\columnwidth} \vspace*{0mm}
\caption{Combined weight simulation at $T=300\,$K: Probability
densities with respect to the distances $d_1$ and $d_2$.
\label{fig_hd1_d2} }
\end{center} \end{figure}

In figure~\ref{fig_hd1_d2} we show the probability densities of 
this simulation with respect to the distances from our reference 
configurations. They are no longer flat, but a satisfactory 
coverage in the variables $d_1$ and $d_2$ is still achieved.
Note that both probability densities have peaks at 
$d=6.62$, which is the distance between configurations 1 and 2.
This implies that both reference configurations have been
visited with high probability.

\subsection{Physics results} \label{phy_res}

We would like to analyze the transitions between our two 
reference configurations in some detail. For this purpose we
use the rms distance, which is defined by
\begin{equation} \label{d_rms}
d_{\rm rms} = \min \left[ \sqrt{{1\over N}\sum_{i=1}^N
(\vec{x}_i-\vec{x}_i^{\,j})^2 } \right]~,
\end{equation}
where $N$ is the number of atoms, $\{\vec{x}^{\,j}_i\}$ are the 
coordinates of the reference configuration $j$, and the 
minimization is over the translations and
rotations of the coordinates of the configuration $\{\vec{x}_i\}$.

\begin{figure}[t] \begin{center}
\epsfig{figure=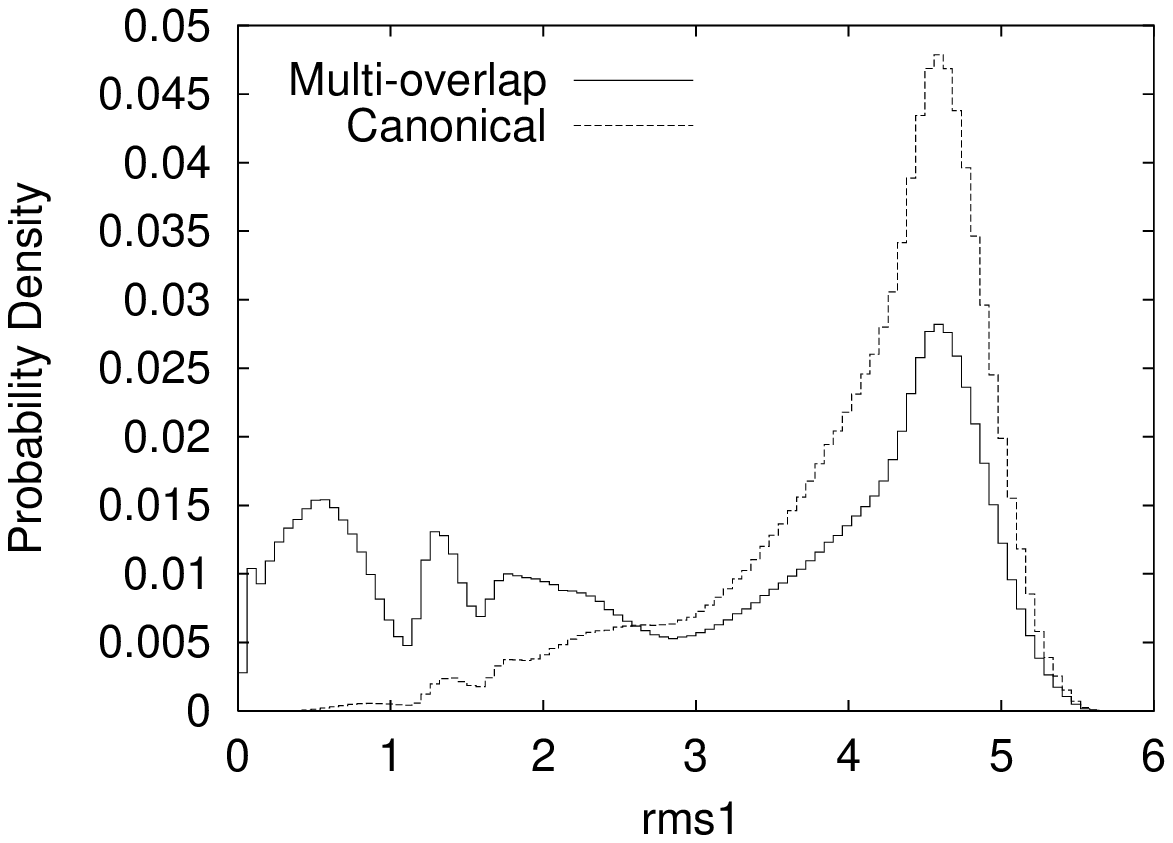,width=\columnwidth} \vspace*{0mm}
\caption{Probability density of the rms distance from the 
multi-overlap simulation at $T=400\,$K of figure~\ref{fig_hist1},
and its canonically re-weighted probability density.
The abscissa is the rms distance (\AA) in Eq.~(\ref{d_rms})
from the reference configuration 1.
\label{fig_rms1} }
\end{center} \end{figure}

The distance~(\ref{dihedral_d}) and the rms distance~(\ref{d_rms})
are quite distinct. The reason is that a change of a single 
dihedral angle in the central parts of the molecule can cause a 
large deviation in the rms distance. Although the two 
configurations are then close-by from the point of view 
of the MC algorithm, physically they are rather far apart, as
the similarity of the three-dimensional structures is 
governed by the rms distance. Therefore,
the rms distance distribution deviates considerably from the
dihedral distance distribution. We illustrate this by
plotting in figure~\ref{fig_rms1} the rms probability density 
of the 400$\,$K simulation for which the dihedral distance 
probability density is shown in figure~\ref{fig_hist1}. 

\begin{figure}[t] \begin{center}
\epsfig{figure=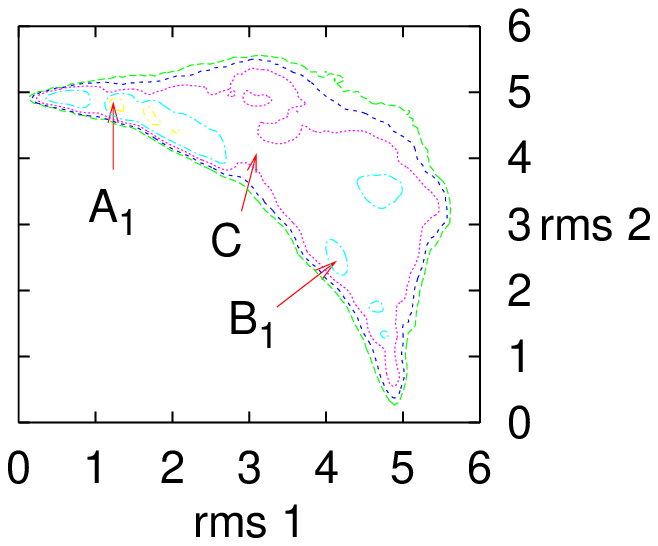,width=9.5cm} \vspace*{0mm}
\caption{Free-energy landscape
at $T=250$ K with respect to rms
distances (\AA) from the two reference configurations,
$F(rms1,rms2)$.
Contour lines are drawn every $2 k_B T$.
The labels A$_1$ and B$_1$ indicate the positions for the local-minimum
states at $T=250$ K that originate from the reference configuration 1
and the reference configuration 2, respectively.  The label C 
stands for the saddle point that corresponds to
the transition state. 
\label{fig_F_rms} }
\end{center} \end{figure}

We now analyze the free-energy landscape~\cite{HOO99} 
from the results of our simulation 
with combined weights at $300\,$K in some detail. 
We study the landscape with respect to some
reaction coordinates (and hence it should be called
the potential of mean force).
In order to study the transition states between reference
configurations 1 and 2, we first plotted the free-energy
landscape with respect to the distances $d_1$ and $d_2$.
However, we did not observe any transition saddle point.
A satisfactory
analysis of the saddle point becomes possible when
the rms distance (instead of the dihedral distance) is used.
Figure~\ref{fig_F_rms} shows contour lines of the free 
energy re-weighted to $T=250$ K, which is close to
the folding temperature~(\ref{Ts}).
Here, the free energy $F(rms1,rms2)$ is defined by
\begin{equation} \label{FE}
F(rms1,rms2)=-k_B T \ln P(rms1,rms2)~,
\end{equation}
where $rms1$ and $rms2$ are the rms distances defined in
(\ref{d_rms}) from the reference
configuration 1 and the reference configuration 2, respectively,
and $P(rms1,rms2)$ is the (reweighted) probability at $T=250$ K
to find the peptide with values $rms1,rms2$.
The probability was calculated from the two-dimensional
histogram of bin size 0.06 \AA $\times$ 0.06 \AA.
The contour lines were plotted every $2 k_B T$ ($=0.99$ kcal/mol
for $T=250$ K).

Note that the reference configurations 1 and 2, which are respectively
located at $(rms1,rms2)=(0,4.95)$ and $(4.95,0)$, are not
local minima in free energy at the finite temperature ($T=250$ K) 
because of the entropy contributions.
The corresponding local-minimum states at A$_1$ and B$_1$
still have the characteristics of the reference configurations
in that they have backbone hydrogen bonds between Gly-2 and
Met-5 and between Tyr-1 and Phe-4, respectively.
We remark that we observe in figure~\ref{fig_F_rms} another
well-defined local minimum state around
$(rms1,rms2)=(4.7,3.5)$.  This state can also be considered to
correspond to configuration 2 because we again observe the
backbone hydrogen bond between Tyr-1 and Phe-4.  The side-chain
structures are, however, more deviated from configuration 2 
than B$_1$, resulting in a larger value of $rms2$.

\begin{figure}[t] \begin{center}
\epsfig{figure=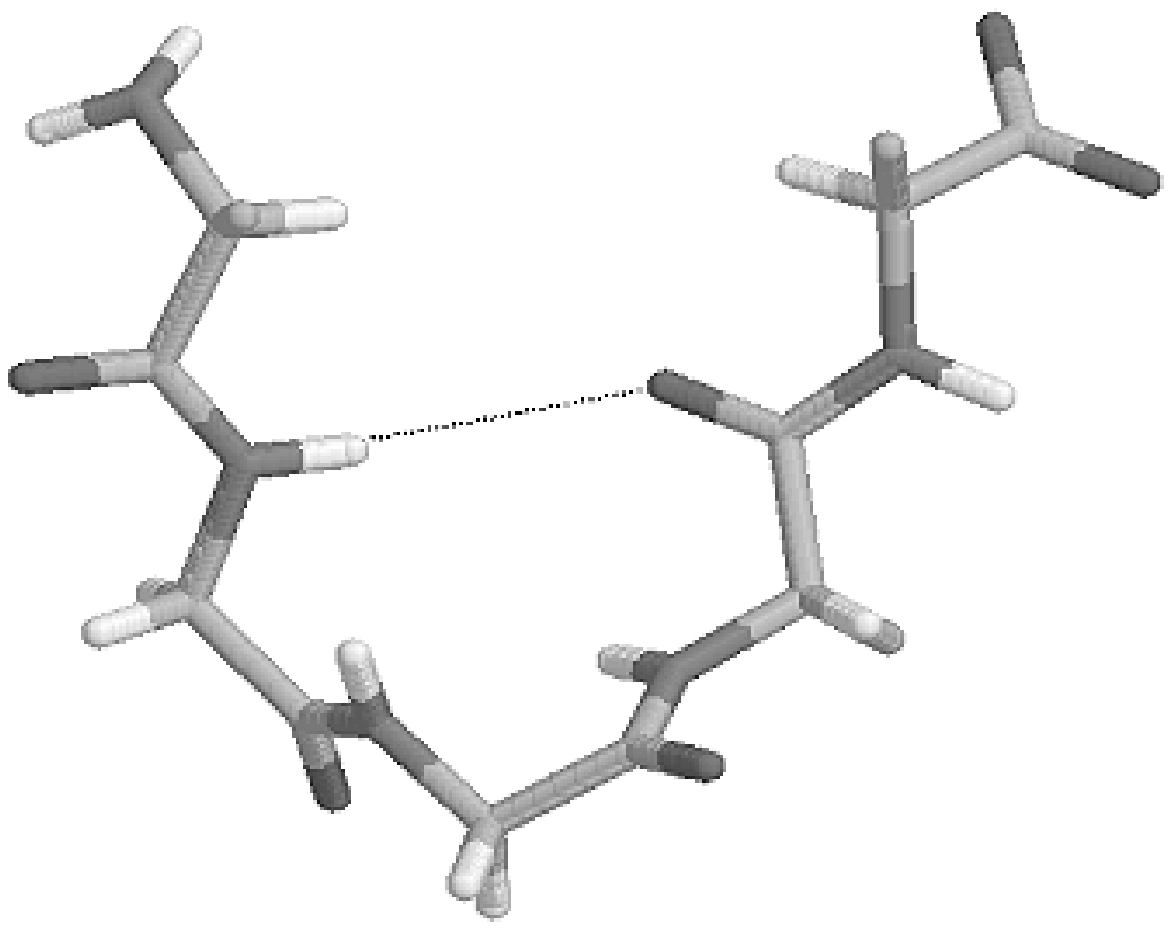,width=7.5cm} \vspace*{5mm}
\caption{The transition state between reference configurations
1 and 2.  See the caption of
figure 1 for details. \label{fig_C} }
\end{center} \end{figure}

The transition state C in figure~\ref{fig_F_rms} should have
intermediate structure between configurations 1 and 2.
In figure~\ref{fig_C} we show a typical backbone structure of
this transition state.
We see the backbone hydrogen bond between Gly-2 and Phe-4.
This is precisely the expected intermediate structure between
configurations 1 and 2, because going from configuration 1 to
configuration 2 we can follow the backbone hydrogen-bond
rearrangements:  The hydrogen bond between
Gly-2 and Met-5 of configuration 1 is broken, Gly-2 forms a hydrogen 
bond with Phe-4 (the
transition state), this new hydrogen bond is broken, and finally
Phe-4 forms a hydrogen bond with Tyr-1 (configuration 2).

It is interesting to see in figure~\ref{fig_F_rms} that there
is only one saddle point in the free-energy landscape
that connects configurations 1 and 2.
Hence, the transition between configurations 1 and 2
always passes through the state C.

In Ref.~\cite{MHO98} the low-energy conformations of Met-enkephalin
were studied in detail and they were classified into several 
groups of similar structures based on the pattern of backbone
hydorgen bonds.  It was found there that below
$T=300$ K there are two dominant groups, which correspond to
configurations 1 and 2 in the present article.
Although much less conspicuous, the third most populated
structure is indeed the group that is identified to be the 
transition state in the present work.

\begin{figure}[t] \begin{center}
\epsfig{figure=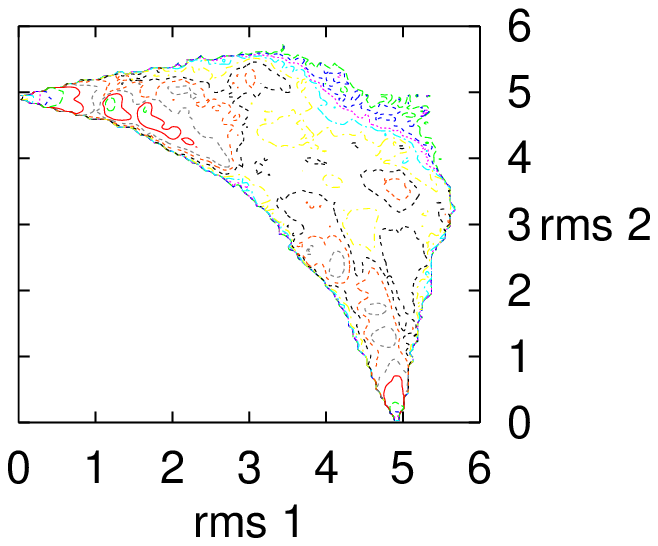,width=9.5cm} \vspace*{0mm}
\caption{Internal energy landscape
at $T=250$ K with respect to rms
distances (\AA) from the two reference configurations, 
$U(rms1,rms2)$.
Contour lines are drawn every $2 k_B T$.
\label{fig_U_rms} }
\end{center} \end{figure}

\begin{figure}[htb] \begin{center}
\epsfig{figure=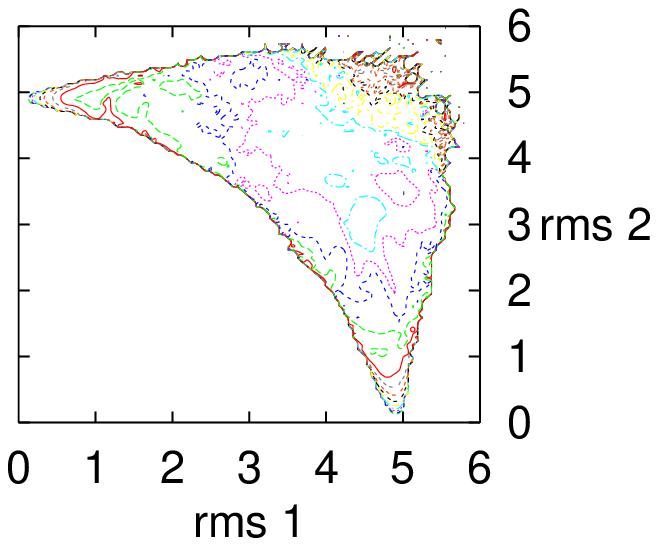,width=9.5cm} \vspace*{0mm}
\caption{Entropy landscape at $T=250$ K with respect to rms
distances (\AA) from the two reference configurations,
$-T S(rms1,rms2)$.
Contour lines are drawn every $2 k_B T$.
\label{fig_TS_rms} }
\end{center} \end{figure}

In figures~\ref{fig_U_rms} and~\ref{fig_TS_rms}
we show the internal energy landscape 
and the entropy landscape at $T=250$ K, respectively.
Here, the internal energy $U$ is defined by the (reweighted) average
ECEPP/2 potential energy: 
\begin{equation} \label{U}
U(rms1,rms2) = <E(rms1,rms2)>~.
\end{equation}
Here, the average was again calculated from the two-dimensional
histogram of bin size 0.06 \AA $\times$ 0.06 \AA.
The entropy $S$ was then calculated by
\begin{equation} \label{TS}
S(rms1,rms2) = \frac{1}{T} \left[U(rms1,rms2) - F(rms1,rms2)\right]~.
\end{equation}
The landscape in figure~\ref{fig_TS_rms} is actually
$-T S(rms1,rms2)$.

Both internal energy and entropy landscapes are more rugged
than free energy landscape (we observe much more number
of contour lines in 
figures~\ref{fig_U_rms} and ~\ref{fig_TS_rms}
than in figure~\ref{fig_F_rms}).
The internal energy has clear local minima at the points
$(rms1,rms2)=(0,4.95)$ and $(4.95,0)$, 
which respectively correspond to configurations 1 and 2,
while the entropy landscape has local maxima at these points.
These two terms tend to cancel each other, and the free energy
landscape is smoothed out.

In table~\ref{tab_FE} we list the numerical values of the
free energy, internal energy, and entropy multiplied by
temperature at the two local-minimum states (A$_1$ and B$_1$
in figure~\ref{fig_F_rms}) and the transition state
(C in figure~\ref{fig_F_rms}).  The internal energy
is just the average of the ECEPP/2 potential energy (without 
any shift of zero point).  The free energy was normalized
so that the value at A$_1$ is zero.  The values at the
coordinates of reference configurations
1 and 2, which are respectively referred to as A$_0$ and B$_0$ in the table,
are also listed.

\begin{table}[ht]
\caption{Free energy, internal energy, entropy multiplied by temperature
at $T=250$ K (all in kcal/mol) at the two local-minimum states (A$_1$ and B$_1$)
and the transition state (C) in figure~\ref{fig_F_rms}.
The values at the coordinates of reference configurations 1 and 2, 
which are respectively
referred to as A$_0$ and B$_0$, are also listed.
The rms distances are in \AA.
\label{tab_FE} } 
\vspace{2mm}
\centering
\begin{tabular}{|c|c|c|c|}
Coordinate (rms1,rms2) & $F$ & $U$ & $-TS$ \\ \hline
A$_1$ (1.23, 4.83) & 0 & $-5.4$ & 5.4 \\ \hline
B$_1$ (4.17, 2.43) & 1.0 & $-3.5$ & 4.5 \\ \hline
C (3.09, 4.05) & 2.2 & $-0.8$ & 3.0 \\ \hline
A$_0$ (0.03, 4.95) & 15 & $-10.5$ & 26 \\ \hline
B$_0$ (4.95, 0.03) & 20 & $-8.1$ & 28 \\ 
\end{tabular}
\end{table}

Among the five points, A$_0$ and B$_0$ are unfavored in free energy
mainly due to the large entropy effects, although they are
energetically most favored.  This means that at
this temperature the exact conformations of 
the reference configurations 1 and 2 are not populated much.
The relevant states are rather A$_1$, B$_1$, and C. 
The state A$_1$ can be considered to be ``deformed'' configuration 1,
and B$_1$ deformed configuration 2 due to the entropy effects, whereas
C is the transition state between A$_1$ and B$_1$.
Among these three points, the free energy $F$ and
the internal energy $U$ are the lowest at A$_1$, while the
entropy contribution $-TS$ is the lowest at C.
The free energy difference $\Delta F$, internal energy difference
$\Delta U$, and entropy contribution difference $-T \Delta S$
are 1.0 kcal/mol, 1.9 kcal/mol, and $-0.9$ kcal/mol between B$_1$ and A$_1$,
2.2 kcal/mol, 4.6 kcal/mol, and $-2.4$ kcal/mol between C and A$_1$,
and 1.2 kcal/mol, 2.7 kcal/mol, and $-1.5$ kcal/mol between C and B$_1$.
Hence, the internal energy contribution and the entropy contribution to
free energy are opposite in sign and the magnitude of the
former is roughly twice as that of the latter at this temperature.

\section{Summary and Conclusions} \label{sec_sum}

We have outlined an approach to perform MC simulations which yield 
the free-energy distribution between two reference configurations. 
The multi-overlap weights for this purpose were obtained by a novel, 
iterative process. The main point of this iterative process is not 
that it is supposed to be more efficient than the recursion that
was used in the multi-self-overlap simulations of 
Ref.\cite{BeBiJa00}, but that it is an entirely independent
approach, which starts from an analytically controlled limit.
Recursions like the one used in~\cite{BeBiJa00} are not 
``foolproof''. For instance, while most of the spin glass 
replica in Ref.\cite{BeBiJa00} were well-behaved, a few did not
complete their recursion after more than an entire year of single
processor CPU time. Similar situations could be encountered in
all-atom simulations of larger peptides, where the normal 
multicanonical weight recursion as well as similar multi-overlap
weight recursion could fail. The present method provides then an 
alternative, approaching the physical region from a different limit.

Noticeable, our multi-overlap approach is well-suited to be combined
with a recently introduced, biased Metropolis sampling~\cite{Be03}.
Namely, the required configurations at higher temperatures are as
well necessary for our particular multi-overlap recursion, so that
no extra simulations are required in this respect.

On the physical side, we have found that entropy effects are 
rather important for a small peptide. 
The effects of entropy on the folding of real proteins in
realistic solvent have yet to be studied in detail.

We have also performed the analysis of this paper for Met-enkephalin
with variable $\omega$ angles and, in particular, simulated 
with combined weights at a number of temperatures. The
results found are quite similar to those reported in this paper. 
In future work we intend to analyze
the transition between reference configuration for larger
systems of actual interest like $\beta$-lactoglobulin. 

\acknowledgments
We are grateful for the financial support from the Joint Studies 
Program of the Institute for Molecular Science (IMS). One of the 
authors (B.B.) would like to thank the IMS faculty and staff for 
their kind hospitality during his stay in spring 2002. In part, 
this work was supported by grants from the US Department of 
Energy under contract DE-FG02-97ER40608 (for B.B.), from the 
Research Fellowships of the Japan Society for the Promotion of 
Science for Young Scientists (for H.N.) and from the Research 
for the Future Program of the Japan Society for the Promotion 
of Science (JSPS-RFTF98P01101) (for Y.O.).

\end{document}